\documentclass{article}

\usepackage{arxiv}

\usepackage{cite}
\usepackage{amsmath,amssymb,amsfonts}
\usepackage{algorithmic}
\usepackage{graphicx}
\usepackage{textcomp}
\usepackage{xcolor}
\usepackage{balance}

\usepackage{paralist} 
\usepackage{lscape}
\usepackage{multirow}
\usepackage{hhline}
\usepackage{afterpage}
\usepackage{units}
\usepackage{float}
\usepackage{enumitem}
\usepackage{xspace}
\usepackage{url}
\usepackage{subcaption}

\def\BibTeX{{\rm B\kern-.05em{\sc i\kern-.025em b}\kern-.08em
    T\kern-.1667em\lower.7ex\hbox{E}\kern-.125emX}}
\begin{document}

\title{Fast Record Linkage for Company Entities}

\author{
Thomas Gschwind, Christoph Miksovic,
	Julian Minder, Katsiaryna Mirylenka, Paolo Scotton \\
\textit{IBM Research -- Zurich} \\
R\"uschlikon, Switzerland \\
\{thg, cmi, jmd, kmi, psc\}@zurich.ibm.com
}

\maketitle

\begin{abstract}
Record linkage is an essential part of nearly all real-world systems that consume structured and unstructured data coming from different sources. 
Typically no common key is available for connecting records. 
Massive data cleaning and data integration processes often have to be completed before any data analytics and further processing can be performed.
Although record linkage is frequently regarded as a somewhat tedious but necessary step, it reveals valuable insights into the data at hand.
These insights guide further analytic approaches to the data and support data visualization.

In this work we focus on company entity matching, where company name, location and industry are taken into account. 
Our contribution is an end-to-end, highly scalable, enterprise-grade system that uses rule-based linkage algorithms extended with a machine learning approach to account for short company names. 
Linkage time is greatly reduced by efficient decomposition of the search space using MinHash. High linkage accuracy is achieved by the proposed thorough scoring process of the matching candidates.

Based on real-world ground truth datasets, we show that our approach reaches a recall of 91\% compared to 73\% for baseline approaches. These results are achieved while scaling linearly with the number of nodes used in the system.
\end{abstract}


\section{Introduction}
\label{sec:introduction}

Enterprise artificial intelligence applications require the integration of many data sources. In such applications, one of the most important entity attributes to be linked is often the company name. It acts as a ``primary key'' across multiple datasets such as company descriptions, marketing intelligence databases, ledger databases, or stock market related data. The technique used to preform such a linkage is commonly referred to as \emph{record linkage} or \emph{entity matching}. 

Record linkage (RL) has been extensively studied in recent decades. RL is in charge of joining various representations of the same entity (e.g., a company, an organization, a product, etc.) residing in structured records coming from different datasets~\cite{MirylenkaSMBA2019}. It was formalized by Fellegi and Sunter in 1969~\cite{Fellegi:RLS}, and the tutorial by Lise Getoor~\cite{Getoor:tutorial} provides an excellent overview of use cases and techniques. Essentially, RL has been used to link entities from different sets or to deduplicate/canonize entities within a given set. To this extent, several approaches have been envisaged ranging from feature matching or rule-based to machine learning approaches. 

Typically, RL is performed in batch mode to link a large number of entities between two or more databases~\cite{Getoor:tutorial}. A challenge of enterprise applications is the ever-increasing amount of unstructured data such as news, blogs and social media content to  be integrated with enterprise data. As a consequence, RL has to be performed between structured records and unstructured documents. This large amount of data may flow in streams for rapid consumption and analysis by enterprise systems. For example, if we consider the analysis of news articles, it is not uncommon to have to process about one million news articles per day. If we consider even only one company name per news, we will have to perform at least ten linkages per second in order to preserve the timeliness of the overall application. Therefore, RL needs to be executed on the fly and with stringent time constraints. 

In this work we assume that the entities to be linked in unstructured data are identified as mentions by a ready-to-use named entity recognition (NER) module such as~\cite{NLU}. 
These entities are then passed to RL in a structured fashion in the form of a record containing attributes that, for example, represent company names, locations, industries and others. RL is in charge of linking this record against one or multiple reference datasets. 

Reference datasets are typically  large and  contain several tens (or even hundreds) of millions of records. When matching an incoming record, it would be prohibitive to perform an extensive comparison to each individual entry in the reference dataset. The most commonly used approach~\cite{Getoor:tutorial} is to decompose the reference dataset into blocks by means of a locality-sensitive hashing (LSH). 

The main contributions of this work are:
\begin{compactenum}
	\item We propose an end-to-end RL system that is highly scalable and provides an enterprise-grade RL service.	
	\item We study scoring functions for various attribute types. We propose a scoring technique for company names, taking into account their specific properties. Also, we propose a hierarchical scoring tree that allows the efficient and flexible implementation of multi-criteria scoring.
	\item We study and apply automatic short company name extraction based on conditional random fields as a sequential classification method. Short names are treated as one of the important features of a company entity. 
	\item We evaluate different LSH configurations as well as our system based on two real-world ground truth datasets. 
\end{compactenum}



\section{Background}
\label{sec:background}

Various record linkage systems have been proposed in recent decades~\cite{Gu:RL, Konda:Magellan, LosterZNMT17, BarbosaCDMPQSS18, KwashieLLLSY19}. As mentioned in the introduction, they can usually be divided into rule-based and machine learning-based systems. We will discuss a couple of these systems that are most relevant to our current and future work.
Konda \emph{et al.}~\cite{Konda:Magellan} have proposed a system to perform RL on a variety of entity types, providing a great flexibility in defining the linkage workflow. This system allows the user to select the various algorithms being used at various stages of the linkage process. Despite its flexibility, this approach does not address the performance problem at the center of the class of applications we are addressing. 
The {\it Certus} system proposed in~\cite{KwashieLLLSY19} exploits graph differential dependencies for the RL task. Even though there is no need for an expert to create these graphs manually, an essential amount of training data is still needed to leverage the graphs automatically. However, we cannot apply such techniques as we consider cases where the amount of training data is very limited.

\textbf{Locality-Sensitive Hashing Methods.}
In the domain of RL, locality-sensitive hashing (LSH) methods are generally used to provide entities with signatures in such a way that similar entities have identical signatures with high probability~\cite{indyk:ANN}. These signatures are commonly referred to as \emph{blocking keys}, which denote \emph{blocks}. Blocks are used to limit the number of comparisons needed during the scoring phase, where candidate entities are compared in detail. 



Typical LSH algorithms are \emph{MinHash}~\cite{Broder:MH, Broder:Resemblance}~\cite{Ullman:MMD} and \emph{SimHash}~\cite{Charikar:SET,Charikar:SH}. MinHash can be parametrized by decomposing the hashing functions in \emph{bands} and \emph{rows}~\cite{Ullman:MMD}. 
The row-band parameter settings for a desired minimal similarity threshold can be determined by the ``S-curve'' calculation as described in the MinHash literature.
In our current setup, we chose MinHash in order to leverage this explicit computation of the parameters. Moreover it can be tuned for high recall (i.e., to maximize the number of relevant entities), which is a prime requirement for RL.

\textbf{Machine Learning for RL.}
A first approach to use machine learning techniques for record linkage was proposed in 2003 by Elfeky et al.~\cite{Elfeky:RLMachineLearning}.
A trained classifier approach is compared to unsupervised clustering and to probabilistic approaches. Although the trained classifier outperforms the other approaches, the authors emphasize the difficulty of obtaining training data. More recent studies  assess the applicability of neural networks to record linkage, for example~\cite{Gottapu:CNNRLS, Mudgal:DLforRLS}. In particular, Mudgal \emph{et al.}~\cite{Mudgal:DLforRLS} show that, compared to ``classical'' approaches, deep learning brings significant advantages for unstructured and noisy data although it achieves only marginal improvements for structured data.

The major limitation to using machine learning techniques for record linkage is the difficulty of finding sufficient annotated training data. This is especially true with company names. Moreover, for each new reference dataset introduced in the system, a specific new training dataset must be developed. To alleviate this problem, some promising approaches such as the use of active learning~\cite{Qian:ActiveLearning} have been proposed. However, the application of machine learning techniques to record linkage remains limited at the moment. 

Nevertheless, machine learning can be applied to sub-problems within record linkage. In this work, we propose a novel machine learning-based technique to extract a short name from a conventional company name. Full company names usually contain many accompanying words, e.g. `` Systems, Inc.'' in ``Cisco Systems, Inc.'', that contain additional information about a company's organizational entity type, its location, line of business, size and share in the international market. The accompanying words often vary greatly from one data source to another. For example, some systems will have just ``Cisco'' instead of the conventional name ``Cisco Systems, Inc.''. These are particularly popular in unstructured data sources such as media publications or  financial reports, where many company mentions are aggregated.

Short company names (sometimes also called colloquial or normalized company names) represent the most discriminative substring in a company name string. 
In many cases, when a query company name is very generic, such as ``Cisco'', there might be a large set of valid correct matches in a reference database. In order to retrieve all possible matching candidates, it is important to extract  and compare corresponding short company names. Short names also allow matching candidates to be found when company names have certain variabilities in both one-to-one and one-to-many linking cases.

It has been shown by Loster \emph{et al.}~\cite{LosterZNMT17} that taking short (colloquial) company names into account is greatly  beneficial for company record linkage. However, the company entity matching system described in~\cite{LosterZNMT17} used a manually created short company name corpus, whereas in this work we focus on  automated short name extraction.
In our deployment, the availability of short company names improves both the efficiency and the accuracy of the RL system, as they lead to smaller and more descriptive blocks and help to give more attention or weight to the most discriminative part of a company name.


\section{Record Linkage System}
\label{sec:recordlinakgesystem}

As mentioned above, we consider the problem of RL performed on the fly, i.e. dynamically linking an \emph{incoming record} to records in one or more reference datasets.
A record is defined as a collection of attributes, each of which corresponds to a column in the dataset. Attributes typically include company name, street address, city, postal code, country code, industry, etc. Note that different reference datasets might not contain the same attribute types, and/or attributes might be referenced by different names. For the latter case, we assume that the attribute names are normalized.

\begin{figure*}[htb]
	\centering
	\includegraphics[scale=0.56]{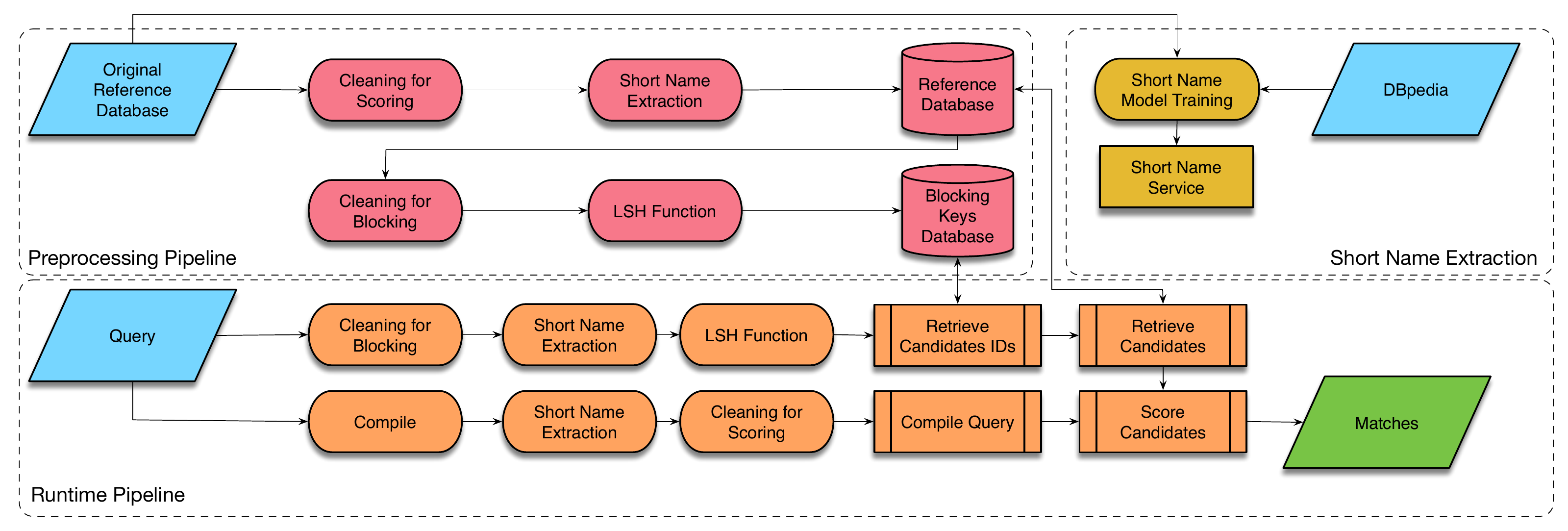}
	\caption{Preprocessing and runtime pipeline.}
	\label{fig:pipeline}
\end{figure*}

The record linkage system essentially comprises three components. One component, \emph{runtime pipeline}, is responsible for matching incoming records against candidate records and returning the best matches. The other two components are used in a an offline mode. These are the \emph{short name extraction system} in charge of training the service to extract short company names and the \emph{preprocessing pipeline}, which prepares the  reference datasets. This architecture is shown in Figure~\ref{fig:pipeline}.

\subsection{Short Name Extraction}
\label{sec:shortnames}

We  use company data from DBpedia~\cite{Auer2007} and from a commercial data source describing company hierarchies as our corpus for short name extraction. The commercial data contains information about companies and their hierarchy, where each company location has its own local identifier. 

{\bf DBpedia.} DBpedia contains some 65~K company entities derived from the English version of Wikipedia.
The company entities contain a name, a label and a homepage of a company. We use all these fields to derive a company short name because, in most of the cases, it appears either in the label or on the homepage of a company. For example, a company called '`Aston Martin Lagonda Limited'' has a label ``Aston Martin''. In this and similar cases based on the handful of heuristically devised rules, we conclude that ``Aston Martin'' is a short name of a company. Similarly, a company called ``Cessna Aircraft Company'' has a homepage \url{http://www.cessna.com/}. Thus, ``cessna'' is used as a short name for training purposes.  After obtaining the short name of a company using its name and homepage, we do additional cleaning in order to exclude the company entity type, such as ``.inc'', ``corp.'' and others. We use the list of business entity types by country provided by Wikipedia~\cite{wikibib}. 
In order to augment the ground truth data with more values, we use two kinds of transformations: from a name to a short name and from a label to a short name. Thus, we make sure that the system will be able to  extract short names correctly not only from the long official name of a company but also from its shorter, sometimes trivial, versions that are often used in free-text sources such as news articles.

\begin{figure}[htb]
	\begin{center}
		\subcaptionbox{DBpedia corpus. \label{fig:histfreq}}[0.475\columnwidth]{\includegraphics[width=0.275\columnwidth]{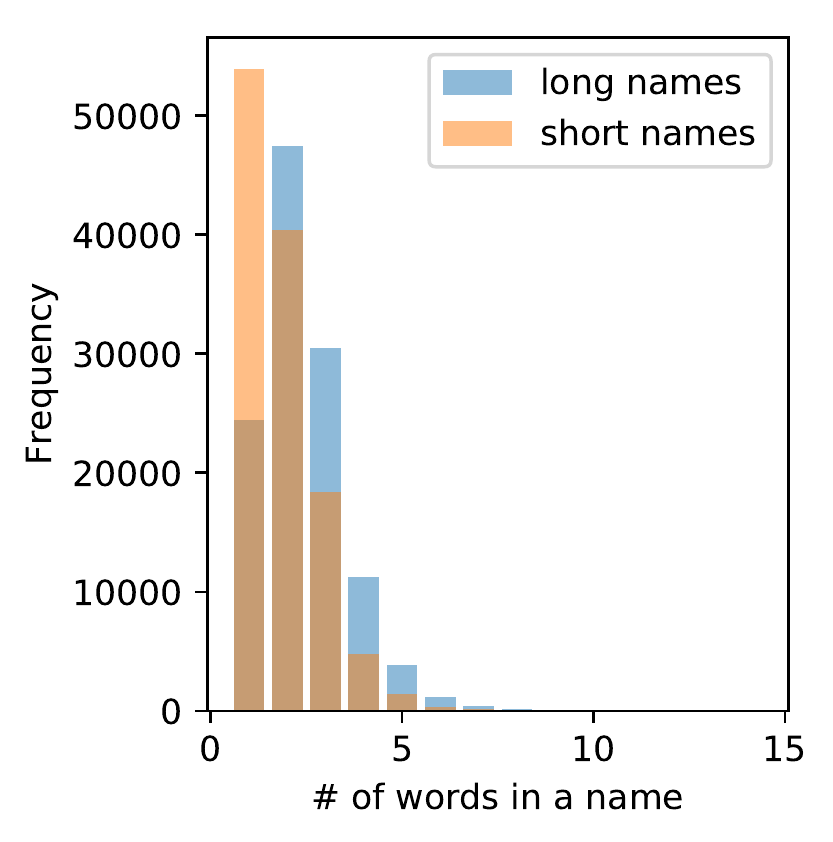}}  
		\subcaptionbox{Commercial company data corpus. \label{fig:histfreq_dnb}}[0.475\columnwidth]{\includegraphics[width=0.275\columnwidth]{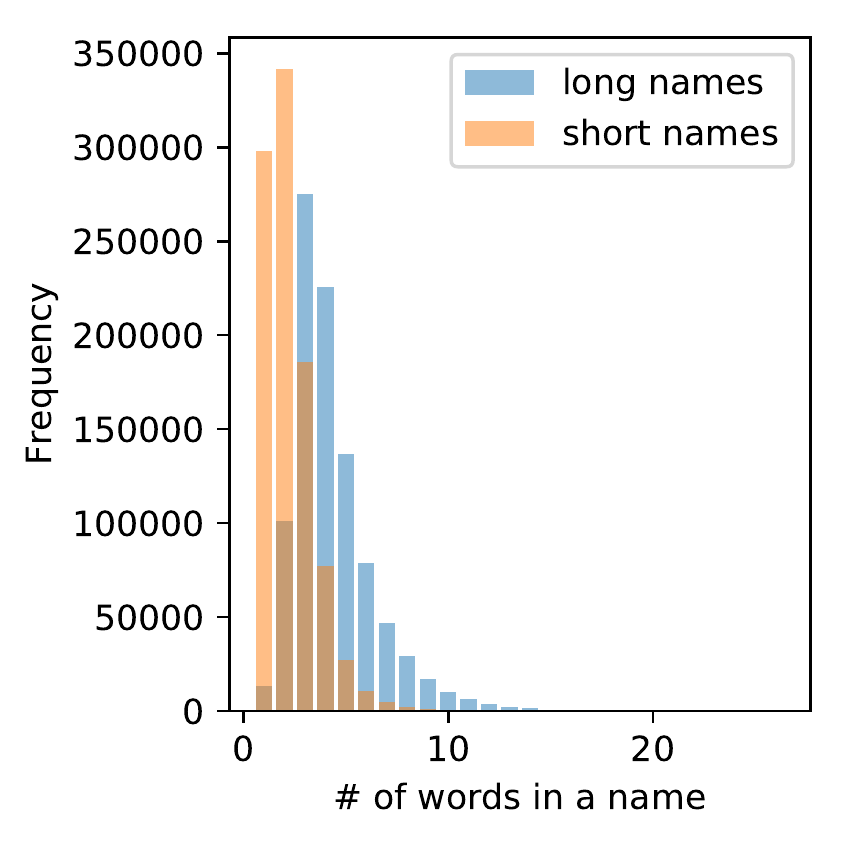}}  
		\caption{Histograms of the number of words in the long and short versions of a company name.}
	\end{center}
\end{figure}

We analyze the number of words in the ground truth data both for long and short versions of the DBpedia company names. 
The distribution of the number of words are shown in Figure~\ref{fig:histfreq}. 
Long company names consist mostly of two or three words up to a maximum of $26$ words. The short names are usually only one or two words. 

{\bf Commercial Company Data.} Another source of training data in our deployment is the commercial company database. This data contains company entities, such as branches, subsidiaries and headquarters, all having individual local and global identifiers. The set of all identifiers associated with a company can be represented hierarchically. Based on these hierarchies, we identify the families of companies which are represented as a tree structure. For each family of companies, we extracted the common tokens of the company names as a short name for the entire family.
After extracting common tokens, similarly to the previous case, additional checks were performed to exclude legal entity types of companies from the token list. The remaining tokens were combined and used as a short name for all the company names in the family. For example, from a family of companies that have two distinct names ``ZUMU HOLDINGS PTY LTD'' and ``ZUMU FOODS PTY LTD'', we extracted ``ZUMU'' to be the representative short name. Given this data source, we were able to extract 950~K of long--short name pairs for training. In total, more than a million pairs of 
names were used as a corpus for automatic extraction of company short names. 

\begin{figure*}[h]
	\begin{center}
		\subcaptionbox{DBpedia corpus. \label{fig:cdf_dbpedia}}[0.45\textwidth]{\includegraphics[width=0.45\textwidth]{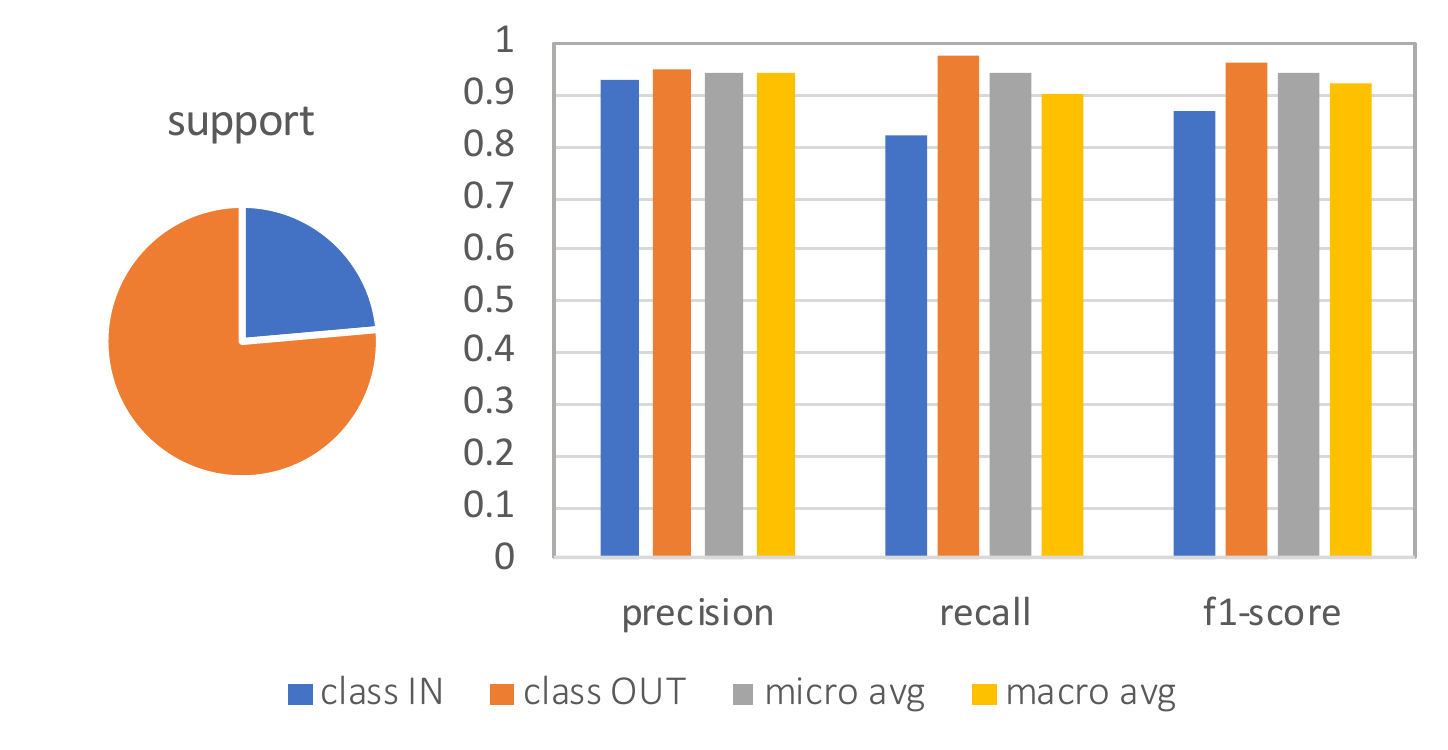}}  
		\subcaptionbox{Aggregated DBpedia and commercial corpus. \label{fig:crf_overall}}[0.45\textwidth]{\includegraphics[width=0.45\textwidth]{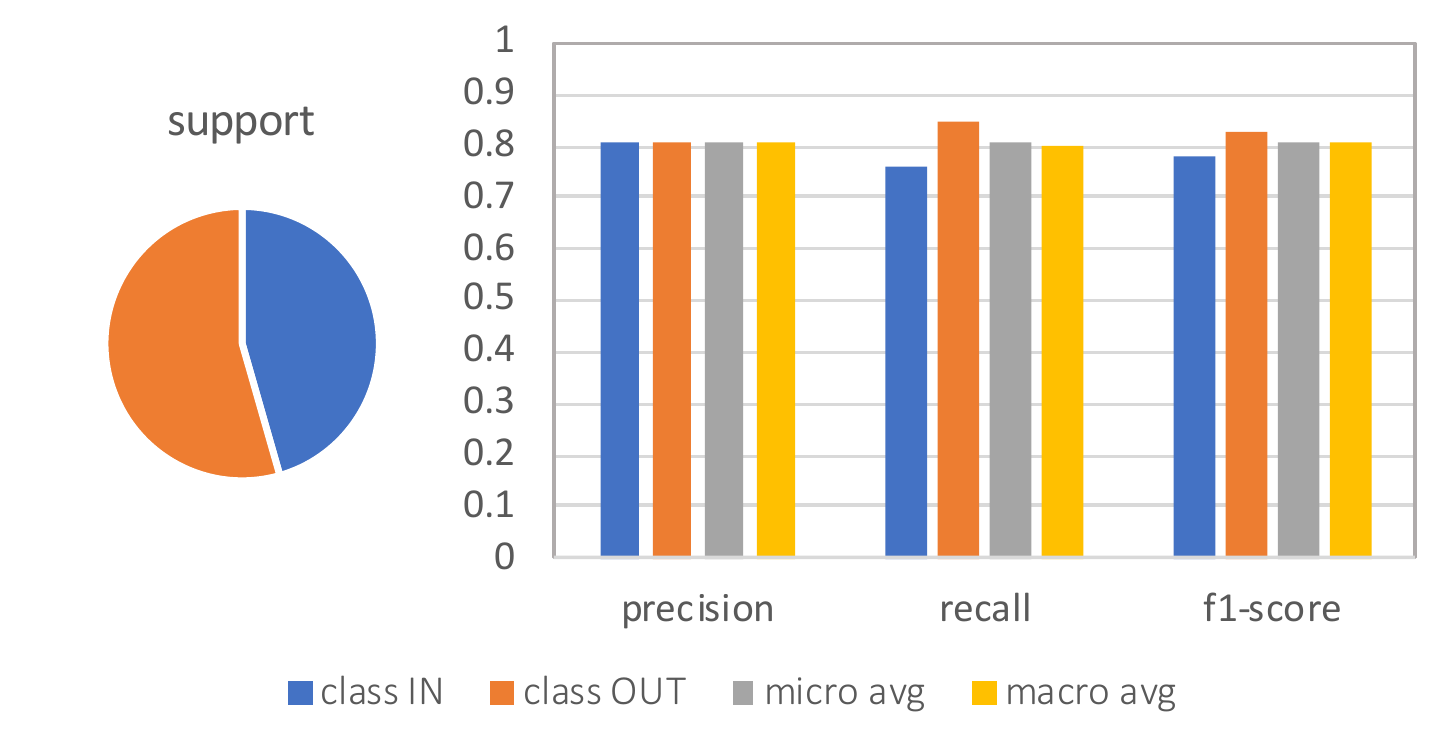}}  
		\caption{CRF performance for company short name extraction.}
		\label{fig:crf_acc}
	\end{center}
\end{figure*}

In the case of commercial company data, the distribution of the number of words within the short and long names is quite different from the DBpedia case (Figure~\ref{fig:histfreq_dnb}). Long names tend to contain more words (3--5 words comprise a large proportion of the probability mass), and short names of only two words are also much more frequent than in the DBpedia case. The task of extracting short names in the case of commercial data is more difficult because the variability of names within the family of companies is greater, and the short name is often the most discriminative part of the name, whereas some other quite discriminative words should be omitted. 
For example, for the following family of companies:
\begin{compactitem}
	\item ``SUNSELEX Verwaltungs GmbH'', 
	\item ``SUNSELEX GmbH solar resources'', 
	\item ``SUNSELEX GmbH solar general constructor''
\end{compactitem}
only ``SUNSELEX'' should be extracted for each of the family members, even though ``solar resources'' and ``solar constructor'' are also quite discriminative words for the corresponding company names. On the other hand, there are cases when quite a ``long'' short company name should be detected. For example, 
``Yunhe County Jincheng''   from the family  ``Yunhe County Jincheng Arts \& Crafts Gifts Factory'' and ``Yunhe County Jincheng Wood Industry Co., Ltd.'', where the number of words that should be kept or omitted in order to obtain the short name is approximately the same. As can be seen from the support pie chart in Figure~\ref{fig:crf_overall}, indeed, for the overall corpus, where the commercial company data portion is dominant, the number of words that should be omitted is slightly greater than the number of words that should be kept in a short name.

We treat the short name learning process as a sequence labeling task, where for each word in a sequence we need to decide whether the word is kept or omitted from a company name. 
Conditional random fields (CRF) ~\cite{LaffertyMP01} is one of the best-performing models applied for sequence labeling~\cite{DBLPHuangXY15}. Currently, the modifications and add-ons of CRF are gaining popularity for complex labeling tasks, such as bidirectional LSTM-CRF models~\cite{DBLPHuangXY15} for part of speech (POS), chunking and named entity recognition (NER) tasks. In our case, we have only two labels: ``IN'' and ``OUT'' to indicate whether the word is included in or omitted from a short name,  respectively. We use the ``just''  CRF classifier, so that only a limited amount of parameters have to be trained.

CRF takes into account the neighborhood of a word to decide on the label and also the predefined set of features for the word and its neighbors. In addition to the usual features used by CRF for NLP sequence labeling tasks, such as the word itself, checks of its capitalization and postfixes, we also include additional features specific to our application:

\begin{compactenum}
	\item frequency rank, which corresponds to the order of the words in a company name according to the overall frequencies found in the training corpus;
	\item normalized frequency or the relative frequency of a word compared to other words in a company name;
	\item absolute frequency.
\end{compactenum}

The feature choice is based on the following. Many company names contain unique words that constitute the short company name. Our hypothesis is that the frequency of a word in a company name has a significant influence on whether the word is included in a short name. We have checked this hypothesis using the DBpedia corpus and, apparently, in $94.6\%$ of the cases, the word with the minimum relative frequency among the words in a company name ends up in the short name of a company. Among the other $5.4\%$ of cases, there are mainly generic company names with relatively small frequency differences among the words, i.e. ``American Fitness Association''. We have used all available data sources in the company domain to compute the frequencies of the words. For example, for a commercial dataset we have used company name, company trade name, city, country and industry type in order to compute overall frequencies.

To evaluate CRF for the task of short name extraction, precision, recall and F1-score are computed separately for ``IN'' and ``OUT'' classes. 
We also present micro and macro averages for each performance measure. The plots for DBpedia corpus and for the aggregated DBpedia and commercial company corpus are shown in Figure~\ref{fig:crf_acc}.

The results demonstrate that CRF is able to distinguish between discriminative and non-discriminative words in a company name as all the performance measures are greater than 0.76 for all the datasets under consideration. Indeed, the task for DBpedia names is easier, and CRF achieves an overall accuracy of approximately 0.9 for both classes. For the larger corpus, the model struggled to reveal all words that should have been included in the short name, providing the recall for ``IN'' class, which is equal to $0.76$. For other performance measures, the values are close to 0.81.

The model is applied to extract the short names in the main record linkage system presented above, with the results in Section~\ref{sec:evaluation}.

It is important for our application to keep the subset of words from the original company name in order to maintain the closest discriminative name representation. 
Thus, we do not produce abbreviations or other name representations that contain only parts of the words in the original company name. In a future work, abbreviations are going to be considered in addition.

\subsection{Preprocessing Pipeline}

The preprocessing pipeline reads records from a given source format,
``cleans'' them, and generates a binary database that supports the
efficient retrieval of the records.  Once the binary database has been
generated, a blocking key database is built by cleaning each
record again and computing for each record a set of blocking key
values corresponding to an LSH function. As discussed in Section~\ref{sec:background}, our implementation uses
MinHash~\cite{Broder:Resemblance,Broder:MH} as its LSH function.  

Generating a MinHash for an entity feature value essentially encompasses the following steps:
\emph{(i)} cleaning, \emph{(ii)} shingling into a set of n-grams, \emph{(iii)} retrieving n-gram vocabulary indices, \emph{(iv)} computing MinHashes for each chosen random hash function, \emph{(v)} grouping MinHashes into bands, \emph{(vi)} hashing band MinHashes, \emph{(vii)} adding band MinHashes to blocking key database.
The band MinHashes are hashed using a general-purpose hash function with an optimal uniform distribution. In our setup, we used the 64-bit Murmurhash3~\cite{Murmurhash3}. 

The blocking key database stores the corresponding record indices  for each blocking key.
We have a ``light''
cleaning operation that is applied to attributes that are used to improve
the scoring performance of candidate records.  
This cleaning operation converts accented characters into their
decomposed form in the Unicode representation~\cite{unicode}, maps
characters to their lowercase equivalent, replaces punctuation
with spaces, and collapses multiple consecutive spaces into
a single space.





To compute the LSH values, attributes are cleaned again
and converted into Jaccard sets using bigrams.  The idea of this
cleaning step is to normalize commonly used notational variations.
For instance, one of our databases contains an entry for
``t{\'e}l{\'e}ski'' whereas another database contains an entry for
``teleski'', both referring to the same entity.  The following table
shows the corresponding Jaccard sets.

\begin{table}[H]
	\centering
	\begin{tabular}{lc}
		Name & Jaccard-Set \\
  		\hline
  		t{\'e}l{\'e}ski & t{\'e}, {\'e}l, l{\'e}, {\'e}s, sk, ki \\
  		teleksi & te, el, le, es, sk, ki \\
	\end{tabular}
\end{table}

Except for \emph{sk} and \emph{ki}, all bigrams are different, yielding a
rather low Jaccard similarity of \(\nicefrac{2}{10}=0.2\). Hence, the probability that these two representations are mapped onto
the same block is low.  Therefore, we apply the following additional
cleaning operations:

\begin{description}
\item[Remove diacritics:] We remove all diacritic marks such as
  accents and umlauts as motivated by the above example.
\item[Remove legal entity types:] We remove the legal entity type of a 	company.  This information is omitted in some databases, which would make it more difficult to match the shortened company name unless the legal entity type is removed.
\item[Merge single characters:] Acronyms are sometimes combined into a
  single word, sometimes separated by dots, and sometimes separated by
  dots and spaces.  This step ensures that bigrams are consistently
  formed for these acronyms.
\item[Merge numbers:] Sometimes numbers are separated using spaces to
  make them more readable.  This operation ensures a consistent number
  representation.
\end{description}

These cleaning operations ensure that records with notational variations are assigned the same blocking key by the LSH function.  Of course, this will generate a number of incorrect matches
that will have to be removed by our subsequent scoring algorithm.

\subsection{Runtime Pipeline}
\label{sec:RuntimeScoring}

The runtime pipeline links queries to the entities stored in the entity database.  It computes the blocking keys and retrieves the corresponding candidate entities. It also transforms the query into a more efficient representation in the form of a tree structure called a \emph{scoring tree}. The scoring tree is evaluated against the candidate entities, and the matching result records are sorted.  Depending on the request, the top-$n$ matching records or all records with a score higher than a given threshold are returned as matches.

The scoring tree uses different scoring algorithms, depending on the type of data to be processed.  If the data describes an address, we use a geographic scoring, whereas if it describes a company name, we use a scoring algorithm tuned for company names.  If multiple types of data are present, the scoring tree combines the scores into a single value.

More formally, the goal of the scoring function $s(.,.)$ is to evaluate the similarity between a query record $R_q$ and a record in a reference dataset $R_r$ such that $s(R_q, R_r) \in [0, 1]$. If $s(R_q, R_r) = 1$ then $R_q$ and $R_r$ are identical.

The query record contains information typically provided by an application. For example, a named entity recognition (NER) component  extracts entities from an unstructured text. By the nature of such a process, the extracted information may vary considerably. Certain attributes might not be recognized, whereas other attributes might be present multiple times. For example, the industry or the street address might not be recognized, but several cities or countries might be present. The query record must be able to  accommodate this variability. Hence, we use a data structure where, except for a company name, other fields are optional and can  have multiple instances. 

The purpose of the scoring function is to compute \emph{(i)} the scores related to individual attributes composing the query and reference records and  \emph{(ii)} the appropriate combination of those scores into a single representative score. The individual characteristics of a record require completely different scoring semantics. Moreover, the combination of individual characteristics follow different rules. For example, when combining scores of company name, address, and industry, a weighted sum of the scores is an appropriate approach. Whereas when combining scores of multiple addresses, a maximum over the individual scores is the right approach.

\subsubsection{Scoring Company Names}
\label{sec:scoringcompanynames}
\label{sec:QueryProcessing:Scoring:CompanyNames} 

In order to score company names, we started with different string similarity functions, such as the Jaccard similarity $j(.,.)$ on which MinHash is based, or the Levenshtein distance $l(.,.)$. However, as we need to obtain a score valued in $[0,1]$ for two strings $s_1$ and $s_2$ representing company names of length $|s_1|$ and $|s_2|$, we use the following formulations:
\begin{equation*}
s_c^{\textnormal{Lev}} = 1 - \frac{l(s_1, s_2)}{|s_1| + |s_2|} \;\;\; and \;\;\; s_c^{\textnormal{Jac}} = j(s_1, s_2),
\end{equation*}
which we call Levenshtein and Jaccard scores, respectively.

A first limitation we observed showed that the Jaccard and Levenshtein scores give too much weight to diacritics. As an example, let us consider a German company name where \emph{\"u} can be written as \emph{ue}. The following table presents the scores with and without the corresponding umlaut:

\begin{table}[H]
	\centering
	\begin{tabular}{lcc}
		& Levenshtein 		& Jaccard \\
		\hline
		\emph{D{\"u}rr} vs. \emph{Durr}		& $0.75$ 			& $0.20$ \\
		\emph{D{\"u}rr} vs. \emph{Duerr} 	& $0.\overline{66}$ & $0.1\overline{66}$  
	\end{tabular}
\end{table}

A na{\"\i}ve approach is simply to remove all the diacritics as part of the cleaning step.  However, there are company names that are only differentiated by the presence of diacritics.  For instance, again from the German language, \emph{ W{\"a}chter}, \emph{Wachter} and \emph{Waechter} all represent existing but different companies. To tackle this problem, we leverage a property of Unicode representation where diacritics are represented as special combining characters. The combining characters are given a lower weight in the scoring process.

Another challenge is to deal with legal entity types of companies such as ``inc.'' or ``ltd.'', which may or may not be included in the company name.  In our initial attempt, we simply removed these legal entity type identifiers.  However, we soon came across companies where the names differ only in the legal entity type but are actually distinct companies. This is one of several occurrences where cleaning had a negative effect on scoring, which confirms the observations made by Randall \emph{et al.}~\cite{randall2013:cleaning}.  Generally, one approach to alleviate the problem related to special mentions (e.g. legal entity types) is to assign them a lower weight in the scoring process. Therefore we adopted the approach of assigning legal entity types the same weight as a single character minus a small value\footnote{The choice of the actual ``small'' value is driven by the Levenshtein function implementation.} of $\epsilon = 1 / 256$. To understand this approach, let us consider three distinct companies \emph{Garage Rex AG}, \emph{Garage Rex GmbH} and  \emph{Garage Rey AG}. The table hereafter shows the Leveshtein score and its modification considering legal entity types:

\begin{table}[H]
	\centering
	\begin{tabular}{lcc}
		&  Levenshtein 					& \begin{tabular}{@{}c@{}} Levenshtein \\ modified \end{tabular} \\ 
		\hline \\
		\begin{tabular}{@{}c@{}}\emph{Garage Rex AG} \\ vs. \\ \emph{Garage Rex GmbH}  \end{tabular} & $1 - \frac{4}{28} = 0.86$    & $1 - \frac{2 - \epsilon}{22} = 0.9092$  \\ \\ 
		\begin{tabular}{@{}c@{}}\emph{Garage Rex AG} \\ vs. \\ \emph{Garage Rey AG}    \end{tabular} & $1 - \frac{2}{26} = 0.92$  	& $1 - \frac{2}{22} = 0.\overline{90}$  
	\end{tabular}
\end{table} 

The fact of subtracting $\epsilon$ allows us to distinguish the case where changes are not in the legal entity type. 

The words in company names can be permuted.  For instance, \emph{IBM Zurich Research Lab} and \emph{IBM Research Zurich} denote the same company. This case is covered by the fact that the Jaccard similarity handles permutations.

In some situations, the city name can be included in the company name. For example, \emph{IBM Research Zurich} is sometimes indicated as \emph{IBM Research} if it is clear from the context that the geographic region is Switzerland.  To handle this, we detect city name mentions in a company name and reduce its weight if the city is in the company's vicinity.  This allows more flexibility with regard to names. To look up city names, we use a fast trie described in the next section.

Additionally, we derive for each company name a \emph{short name} as described in Section~\ref{sec:shortnames}. Words that are part of the short name are weighted higher, namely three times the normal weight. This approach allows us to place more emphasis on the characteristic words of the company compared to other elements present in the name. 

Finally, the company name score is computed as:
\begin{equation*}
s_c (R_q, R_r) = 0.9 \, \max (s_c^{\textnormal{Jac*}}, s_c^{\textnormal{Lev*}}) + 0.1 \, \min(s_c^{\textnormal{Jac*}}, s_c^{\textnormal{Lev*}}),
\end{equation*}
where $s_c^{\textnormal{Lev*}}$ and $s_c^{\textnormal{Jac*}}$ are  the Levenshtein and Jaccard scores, respectively, modified with the considerations above, and applied to the company names in $R_q$ and $R_r$. The rationale behind this choice is that the Jaccard score allows for word permutations, whereas the Levenshtein score relies on the character sequence. 
However, this fact is not well represented by a weighted average.
The reason we do not simply use the maximum between the two
similarities is that, in certain cases, the Jaccard similarity may return a similarity
of 1 for names that are different.  This is to ensure that a match
with a Jaccard similarity of 1 is not chosen coincidentally over a
Levenshtein similarity of 1, which is only possible if the strings
are equal.  The values of 0.9 and 0.1 have been chosen arbitrarily.

We dubbed this measure \emph{RLS}. In Section~\ref{sec:evaluation} we will evaluate this approach against the following strategies:

\begin{description}
	
	\item[Jaccard:] uses the Jaccard score
	(\(s_{\textnormal{Jacc}}\)).
	
	\item[Levenshtein:]  uses the Levenshtein score (\(s_{\textnormal{Lev}}\)).
	
	\item[weighted:] uses the arithmetic mean between the Jaccard and
	Levenhstein scores
	(\(s_{\textnormal{weighted}} =
	\nicefrac{1}{2}(s_{\textnormal{Jacc}}+s_{\textnormal{Lev}})\)).
	
	\item[max-min:] This is the scoring the RLS, except that
	the combining characters, legal entity type, and city optimizations are not
	activated
	(\(s_{\textnormal{maxmin}} =
	0.9\cdot\mbox{max}(s_{\textnormal{Jacc}},s_{\textnormal{Lev}})+0.1\cdot\mbox{min}(s_{\textnormal{Jacc}},s_{\textnormal{Lev}})\)).

\end{description}

\subsubsection{Scoring other attributes}
\label{sec:scoringotherattributes}

Multiple attributes can be taken into consideration for scoring; in this section we briefly describe geolocation and industry scoring.
 
A geographical location is described by means of an address element. This element contains the street address, postal code, city and country code attributes. Each component is scored using a specific algorithm. The street address is currently scored using a tokenized string matching (e.g. Levenshtein tokenized distance~\cite{mirylenka2018}). This provides a reasonable measure between street address strings, especially if street number and street name appear in different orders. Postal codes are evaluated according to the number of matching digits or characters. The rationale behind this approach is that, to the best of our knowledge, the vast majority of postal code systems are organized in a hierarchical fashion. However, this scoring can be improved by using a geographic location lookup service. If the GPS location is available in the reference dataset, the city is scored using the Haversine distance~\cite{Inman1835}. To retrieve the GPS location of the city mentioned in the query record, we use a trie data structure, which contains the names and GPS position of some 195,000 cities worldwide obtained from \texttt{geonames.org}~\cite{geonames}.  To evaluate the score, we compute the Haversine distance between cities associated with an exponential decay in $[1,0[$. As a fallback, if the GPS position is not available or the city in the query record cannot be found in the trie, we use the Levenshtein score (described in \ref{sec:scoringcompanynames}) between city names. Finally, the country code is a simple comparison: If the countries match, the score will be 1. If not, the score is 0.

Industries are typically represented by four-digit Standard Industry Classification (SIC) codes~\cite{SIC}. Similar to postal codes, SIC industry codes are also hierarchical: The first two left-most digits represent a ``Major Group'' (e.g. Mining, Manufacturing and others), the following digit is the ``Industrial Group'' and the last digit is the specific industry within the industrial group. When representing an industry, codes of variable length can be used, depending on the level of generality of the representation. To evaluate the industry score, we use a measure similar to the one used for postal codes.

\subsection{Implementation}
The design of our RL system is driven by three main goals: versatility, speed and scalability.

Versatility is given by the generality of the approach. As we have shown in the previous sections, the various components have been designed to be able to accommodate virtually any reference dataset and to perform RL on a large variety of entities. The scoring function set can be extended to other attribute types, e.g. product names, person names, and others. Also, the scoring tree can be adapted to accommodate these new attribute types with appropriate combining functions. The central element of the system is a generic ``linker'', which can easily be configured to load a preprocessed dataset and perform linkage. To maximize performance in terms of speed, the linker has been written in C++ and loads the entity database into memory. Therefore, once the linker is started and initialized, all operations are performed in memory. In addition, the linker uses a multi-threaded approach such that asynchronous RL requests can be processed in a parallel fashion to exploit the cores available on the physical system. Each reference dataset and, therefore, the associated entity databases are loaded in a specific linker.

To ensure scalability, we have adopted a containerized approach; each linker runs in an individual container. In conjunction with a container orchestration system, such as Kubernetes, it is possible to run and dispatch linkers on multiple physical machines. This approach allows  linear scaling with the number of nodes that are added to the cluster as well as the ability to run linkages simultaneously against multiple datasets. Moreover, the overall system is resilient to node failures, which is an important characteristic for an enterprise-grade application.

\section{Evaluation}
\label{sec:evaluation}

We evaluated our RL system over multiple dimensions.  First, we
assessed the parameters under which it yields the best performance,
that is identifying the right tradeoffs between memory usage and
performance as well as among the different scoring strategies
available.

Second, we compare our scoring algorithm to using just standard
scoring algorithms.

Third, we compare our RL system to
two baseline systems: using a simple case insensitive string lookup
of the company name (to identify the number of trivial matches) as
well as using Apache Solr~\cite{ApacheSolr}, a state-of-the-art
distributed indexing system that powers the search and navigation
features of many of the world's largest internet sites.

Fourth, we evaluate the scalability in relation to the number of
parallel clients accessing the system.  

To evaluate these dimensions, we rely on two commercial but publicly
available company databases.  The first company database comprises
150~million records (we also call it the reference database above in the paper) and is used by any company that engages in
government contracts in the United States.  The second is used by
financial analysts and comprises approximately 15~million records.  
Finally, we also use internal accounting data comprising 2~million records.
We denote these company databases as 150m, 15m, and 2m
respectively.

\noindent \textbf{Swiss dataset.}  We randomly selected 450 companies
located in Switzerland from the 15m database and manually matched it
against the 150m database.  The use case behind this dataset is to
identify local companies  with a random mix between small,
medium, and big companies as would be encountered by a user with a
strong local interest.  Switzerland was chosen for two reasons: \emph{(i)}
it has four different official languages allowing us to
assess the system in combination with different languages and \emph{(ii)} our
familiarity with the region was instrumental in correctly identifying
records referring to the same company. 
Company records were divided into the following categories:

\begin{compactdesc}
\item[Matched] In this case, all correct matches are listed.  For
  instance, if 15m database was missing the address, or listed many
  subsidiaries and 150m database only listed the
  headquarters. (296~records, 196~unique records)
\item[Unmatched] If no corresponding company was present in the
  company database. (114~records)
\item[Undecided] In cases where we were unable to decide conclusively
  whether companies are the same or if one of the companies had been
  renamed, e.g. following a merger.  These records were counted
  neither as true positive nor as negative but only as false positive
  if they were matched against a different record. (80~records)
\end{compactdesc}

\noindent \textbf{Accounting dataset.}  We leveraged internally available
financial data that maps accounting company data from the 2m database to
the 150m database.  This dataset consists of 55k records.  As this
linkage was manually performed by domain experts, we can assume that
$>99\%$ of detected links are accurate.  
The use case behind this dataset is
to match accounting data against a reference database as is it usually
performed in large companies.

\noindent \textbf{News dataset.}  The dataset is based on a random selection of
104~current news articles from different sources.  It was
manually curated and lists for each
article the companies that should be found from the reference company database.  
The use case behind this dataset is to mine data about
companies from unstructured data sources.

\subsection{Performance Tuning}
\label{sec:Evaluation:PerformanceTuning}

\begin{figure}
\centering
\includegraphics[width=0.6\columnwidth]{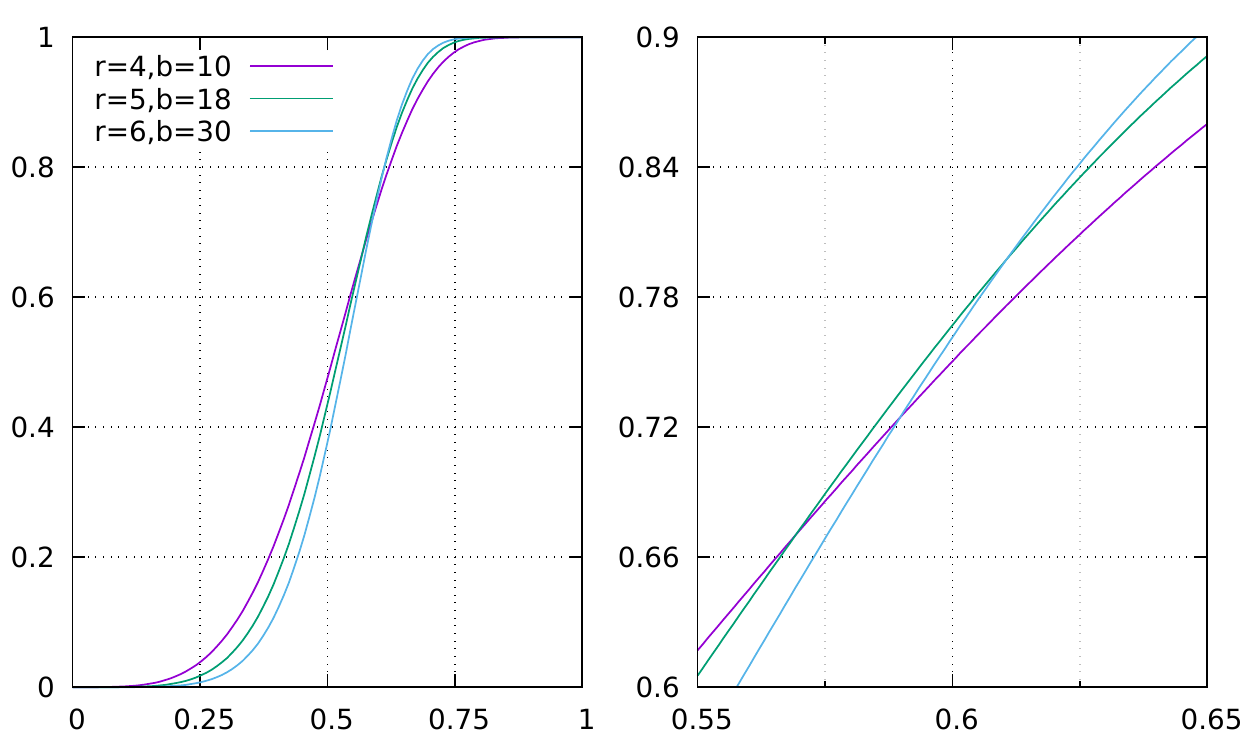}
\caption{S-curves (left: full; right: zoomed; $x$-axis:
  Jaccard similarity; $y$-axis: matching probability)}
\label{fig:scurves}
\end{figure}

In this section, we evaluate different tuning parameters and their
effects on the performance of our RL system.  First, we analyze
different MinHash row-band configurations.  From experience, we
identified that correct matches typically have a Jaccard similarity
 greater than~0.8.  However, some correct matches have a
score as low as~0.6.  Using these numbers, we have chosen three
row-band configurations such that matching records with a Jaccard
similarity of $\ge 0.8$ are matched with a probability $>99\%$ and
those with a similarity of $<0.8$ with a probability $>75\%$.
Considering that correct matches with a score $<0.8$ are rare, the
75\% figure has been arbitrarily chosen as a tradeoff between
performance and matching accuracy.  The row-band configurations are
shown in Figure~\ref{fig:scurves} and Table~\ref{tab:scurves:prob}.


\begin{table}
  \centering
  \begin{tabular}{lrrr}
    minhash & 4/10 & 5/18 & 6/30 \\
    \hline
    $\sigma=0.5$ & 47.5\% & 43.5\% & 37.6\% \\
    $\sigma=0.6$ & 75.0\% & 76.7\% & 76.1\% \\
    $\sigma=0.7$ & 93.5\% & 96.3\% & 97.6\% \\
    $\sigma=0.8$ & 99.4\% & 99.9\% & 99.9\% \\
  \end{tabular}
  \caption{Minhash matching probabilities}
  \label{tab:scurves:prob}
\end{table}

These configurations allow us to evaluate the tradeoff between the
different MinHash configurations.  A higher number of rows and bands
yields a sharper S-curve (described in~\ref{sec:background}).  Hence,
entities with a low score are less probable to be considered a
match. However, this comes at the expense of having to compute more
MinHashes (numerically, $\mbox{rows} \times \mbox{bands}$) as well as
consuming more memory to store the additional bands.

Table~\ref{tab:bandsperformance} shows the results of linking the Swiss dataset  against 150M database.
The recall figures for the different configurations are approximately
87\% ($\pm 0.5\%$).  This is not surprising, considering that the
S-curve was configured to capture all company names with a Jaccard
similarity of $\ge 0.8$ with a $>99\%$~probability.  Memory
consumption almost grows linearly with the number of bands.  The
number of comparisons necessary, however, looks surprising, because
the S-curves are relatively close to each other.  After considering
that the scoring distribution among candidate entities has a heavy
tail distribution with considerably more candidate entities having a
low Jaccard similarity, this difference is easily explained.  Finally,
the time spent computing the $6 \times 30$ MinHashes for each record
to be matched was negligible, which, looking at the table, is due to
the fact that each record must be compared only with 23~thousand to
72~thousand candidate records.

\begin{table}
  \centering
  \begin{tabular}{lrrr}
    MinHash          &    4/10 &    5/18 &    6/30 \\
    \hline
    recall           & 86.67\% & 87.18\% & 87.18\% \\
    database size    & 38.8GiB & 57.6GiB & 99.5GiB \\
    comparisons      &   72.9k &   55.0k &   23.6k \\
  \end{tabular}
  \caption{Memory and performance comparison}
  \label{tab:bandsperformance}
\end{table}

We chose the configuration with 30 bands for our RL system as it provides
the best tradeoff between memory consumption and comparison operations required.


\subsection{Scoring Evaluation}
\label{sec:Evaluation:Scoring}

\begin{figure}
\centering
\includegraphics[width=0.6\columnwidth]{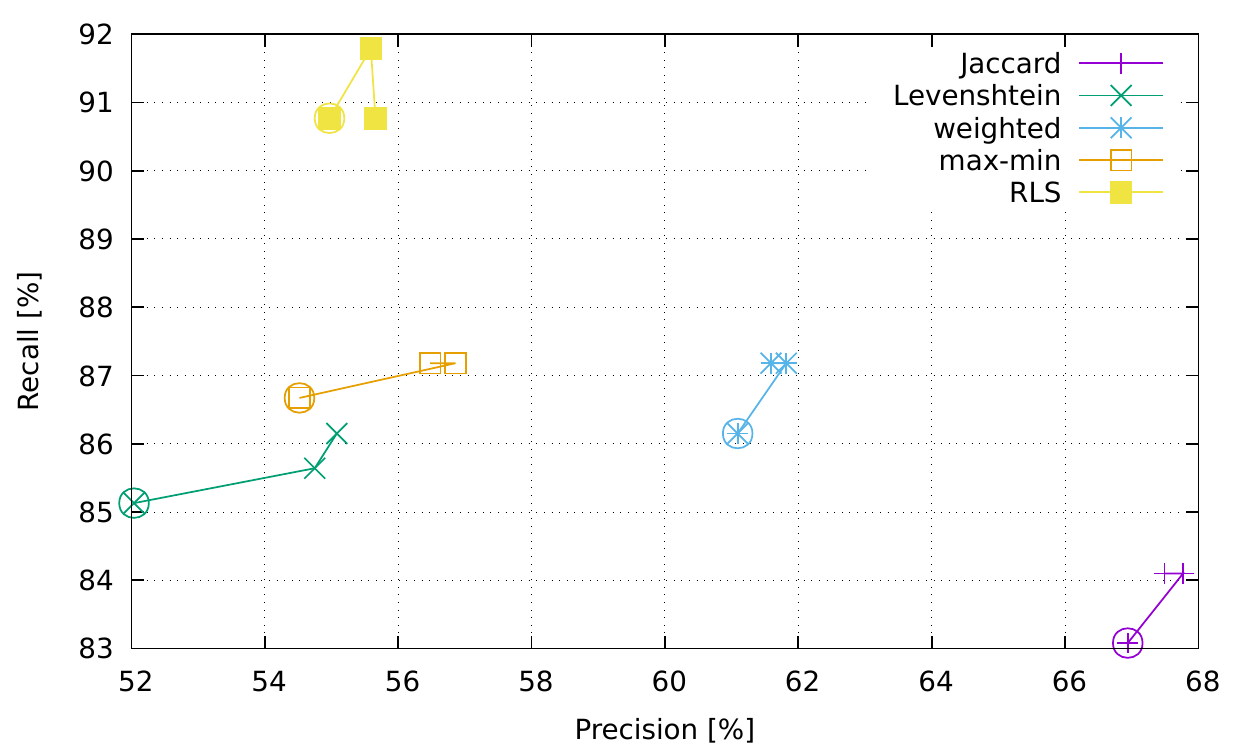}
\caption{Comparison of distance functions for the different MinHash
  configurations: $r=4$, $b=10$ (encircled data point), followed by
  $r=5$, $b=18$ and $r=6$, $b=30$ ($x$-axis: precision [\%]; $y$-axis: recall
  [\%])}
\label{fig:scoring0}
\end{figure}

In Section~\ref{sec:QueryProcessing:Scoring:CompanyNames}, we have
described our algorithm for scoring company
names. Figure~\ref{fig:scoring0} compares the precision and recall
numbers for the different strategies.  The similarity functions are
shown for the row-band configurations discussed previously: $r=4,b=10$
(encircled), followed by $r=5,b=18$ and $r=6,b=30$.  The results are
very similar for the different band configurations, a bit
better for those with higher band numbers, which would be supported by
the fact that the matching probability for similarities $>0.6$, to
include most outliers, is slightly higher for higher band numbers.

The Jaccard similarity has a lower recall than the Levenshtein
similarity because the former is more sensitive to small changes in
the name, as we have seen in the ``\emph{D{\"u}rr} example'' 
in Section~\ref{sec:RuntimeScoring}.
As a consequence, its precision is higher.  The weighted approach lies
somewhere in the middle.

The max-min strategy compared to the weighted strategy yields similar
results in terms of recall but with lower precision. This can be
explained in the case where two company names have a ``high'' Jaccard
score and a ``low'' Levenshtein score. For example, if the Jaccard
score is 1.0 and the Levenshtein score is 0.4, then the arithmetic
mean is 0.7, which is barely above our threshold, whereas max--min
yields a score of $0.94$, which closely resembles the Jaccard
similarity.

The comprehensive RLS approach shows significant improvements in terms
of recall. The precision is similar to the max-min strategy but below
the weighted or Jaccard strategies. This is because it finds matches
for records in the ground truth dataset that have no corresponding
matches in the reference database. In this case it is almost
impossible to discern close matches from non-matches. Considering that
we favor recall over precision, this is an acceptable tradeoff.

\subsection{Matching Accuracy}
\label{sec:Evaluation:MatchingPerformance}

In this section, we evaluate the  matching accuracy -- in
terms of recall and precision -- of the RL system.  We compare to 
a trivial case 
string comparison approach, to identify the
number of trivial matches, as well as to a Solr based approach.

Solr was configured in a manner to allow for flexible search
operations that do not impose rigid restrictions on the type and
structure of the query terms. A default search field based on the
``solr.TextField'' class including standard tokenization and
lowercasing was used to copy all the relevant company attributes into
one multivalued field. This setup allows to submit compact query data,
e.g. company name only, as well as complex query phrases that contain
a company name and arbitrary additional attributes like address, city,
and country.

\begin{table}
  \centering
  \begin{tabular}{lrrr}
	       & Trivial Matches & Solr & RLS \\
\hline
Swiss	   & R:  57\% & R:  72\% & R:  91\% \\
           & P: 100\% & P:  41\% & P:  65\% \\
Accounting & R:  28\% & R:  73\% & R:  89\% \\
           & P: 100\% & P:  73\% & P:  89\% \\
News	   & R:  33\% & R:  42\% & R:  64\% \\
           & P: 37\%  & P:  42\% & P:  58\% \\
  \end{tabular} 
  \caption{Matching Performance of Trivial Matching vs.\ Solr vs.\ RLS}
  \label{tab:evaluation:MatchingPerformance}
\end{table}

The performance results for  all our datasets (Swiss, Accounting, News) are
summarized in Table~\ref{tab:evaluation:MatchingPerformance}.

\textbf{Swiss dataset.} For the first use case, we see that the
trivial matching algorithm is already able to correctly match 57\% of
the records of the Swiss dataset.  We assume that this is because many
companies ensure that their information is correctly stored in the
150m and 15m databases.  Precision is at 100\% because all matched
names have been identified correctly.

Compared to the trivial approach, Solr is able to match another 15\%
of the records, i.e., 35\% of those records not matched by the trivial
approach.  
shortened or extended. 
 RLS is able to match yet another
19\% of the records that have not been matched by Solr, or in other
terms, 79\% of the records not matched by the trivial approach.  The
reason is that RLS is aware of matching semantics of 
different artifacts that compose a company record.  The relatively low
precision is explained by the fact that both, Solr and RLS, try to
find a match for every record.  The precision for Solr is lower as
as it is less specialized for the task of company matching.

\textbf{Accounting dataset.}  The recall values of the accounting
dataset mostly mirrors the Swiss dataset despite storing mostly bigger
and international companies that frequently use English words in their
names.  Interestingly the trivial matching performs much worse.  This
seems to be because the accounting database with 2m records is only
internally available and hence at times uses unofficial name
variations of the company.  These variations are mostly trivial and
hence both, Solr and RLS, perform similarly to the Swiss dataset.

The precision values for Solr and RLS are the same as the recall
values since this dataset only contains records that are present in
both databases and both systems have returned a match for every record
they have received.


\textbf{News dataset.} For this dataset, as discussed previously, we
use Watson NLU to identify company names in news articles. Subsequently
these names are matched to  the 150m database. This task is much
harder as very limited context is available and company names may  
vary substantially. This leads to lower accuracy results. Due to its specialization and extra-processing,
RLS still outperforms Solr. 

It has to be noted that we only used the company name as other
attributes are often not reliable or not present in the unstructured case. For example, an
article may contain several company names, cities and countries and
therefore it can be ambiguous to an automated entity resolution system
which city/country refers to a given company.  


\subsection{Scalability}
\label{sec:Evaluation:Scalability}

Each RLS instance accepts up to 8 parallel client requests.  
A request can contain multiple individual queries that are distributed over 4 threads.
This gives a theoretical maximum of 32 queries being processed in parallel. Each instance runs on a 
server with two Intel~\textsuperscript{\textregistered} 
Xeon~\textsuperscript{\textregistered} CPU E5-2630 2.2GHz 
(total 40 cores) and 400GB of memory. 

\begin{figure}[htbp]
  \centering
  \includegraphics[width=0.6\columnwidth]{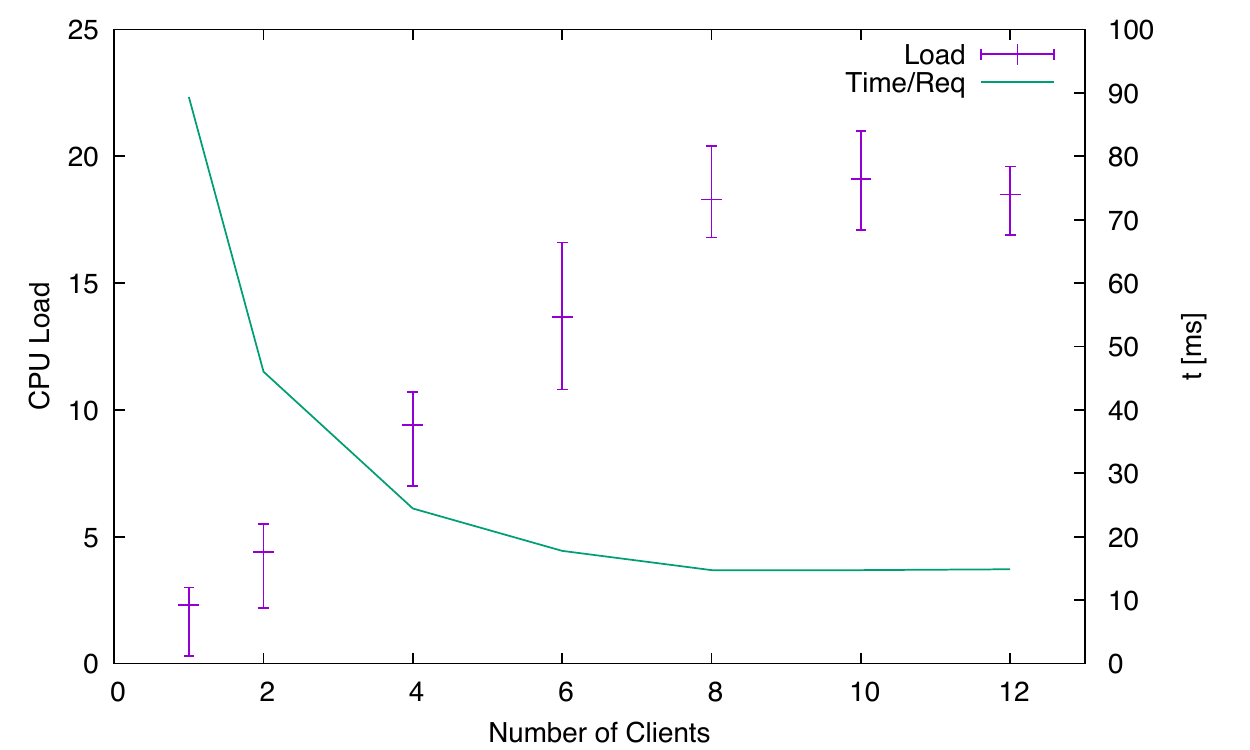}
  \caption{Scalability Analysis of the RL System}
  \label{fig:Evaluation:Scalability}
\end{figure}

We have deployed this service on a node which is part of our Kubernetes cluster.
Requests are issued by a range of one to twelve parallel clients. Each client
sends 10'000, requests each containing  80~queries. The scalability results are shown
in Figure~\ref{fig:Evaluation:Scalability}.  Up to eight clients, 
the CPU load increases almost linearly. We notice that the average
processing time decreases as a benefit of parallel processing. 
It converges to a value of 17~ms per request. 
With more than eight clients, the CPU load and performance gain levels off 
as requests start to be queued.

Scaling requests over multiple nodes is performed by the load balancer
of Kubernetes.  Since each instance keeps its own copy of the reference company
database and hence runs independently, no performance penalties are
incurred by Kubernetes scaling the service over multiple nodes.


\balance

\section{Conclusions and Future Work}
\label{sec:conclusion}

In this work, we presented the fast RL system for company matching. We showed that the proposed system is able to accurately match about 30\% more records compared to the baselines. This improvement is due to two contributions:  \emph{(i)} the introduction of short company name extractions and their use both in the preprocessing phase as well as in the scoring phase and \emph{(ii)} specific improvements of the scoring function, namely taking into account diacritic characters, legal entity type, and the ability to identify geographic locations in a company name.

Additionally, as deployed today in a cluster with three nodes, the system is capable of an aggregated processing time of 17~ms per record, which means that we are able to match approximately 5M records per day. These performance figures scale linearly with the number of nodes, making the system perfectly suited for analyzing high-volume streamed contents.

Our future work will:
\begin{compactitem}[$\bullet$ \leftmargin 0pt]
\item explore the use of other LSH functions such as SimHash~\cite{Charikar:SH} to assess whether our recall values can be improved further,
\item maintain automatic parameter learning and automatic training dataset augmentation,
\item take into account company abbreviations,
\item consider the historical evolution of company names and additional company modeling~\cite{mirylenka2016applicability, mirylenka2016recurrent, MirylenkaSMD19}.
\end{compactitem}

\balance

%

\end{document}